# Security Threats and Their Impact on Blockchain Interoperability: Identification and Countermeasures

Shawn M. Reynolds, Hassan Reza
School of Electrical Engineering and Computer Science
University of North Dakota
Grand Forks, ND 58201, USA
Shawn.reynolds@und.edu; Hassan.reza@und.edu

**Abstract—**Blockchain interoperability enables independent blockchain systems to communicate and exchange assets across heterogeneous networks. A major weakness limiting its effectiveness is the lack of comprehensive security mechanisms. Attacks have already caused hundreds of millions in asset losses, demonstrating the severity of this threat. This paper identifies and counters all known attacks and vulnerabilities facing interoperable blockchain systems. We classify threats into five categories: (1) core blockchain attacks, (2) network attacks, (3) interoperability-specific attacks, (4) social engineering, and (5) code vulnerabilities—particularly in smart contracts. For each threat, we identify effective countermeasures. By cataloging these threats and their solutions, we provide a foundation for designing secure blockchain interoperability solutions.

**Keywords—**blockchain interoperability; security threats; countermeasures; smart contracts; cross-chain attacks

## I. INTRODUCTION

Blockchains (BCs) are decentralized, immutable data structures built on distributed database systems. A BC is a sequence of blocks connected by hash pointers; any modification is detectable because hash values would no longer match [4]. Nodes maintain copies of the BC and reach consistency through consensus, ensuring new blocks are appended only when most nodes agree [4]. Smart contracts extend BC functionality through self-executing programs that automate predefined actions [1].

Unfortunately, BCs cannot natively interoperate with one another, drastically limiting their utility. BC interoperability enables different blockchains to communicate, exchange data, and transfer assets [1]. Solutions enabling these interactions are often called bridges [14]. However, building interoperability on existing, heterogeneous BC technology is challenging, and security is among the most pressing concerns before ever considering interoperability.

Security threats compound in interoperable systems: vulnerabilities exist at the base BC layer, at the network layer, and in the cross-chain protocols themselves [15]. Attacks have already succeeded against real systems—the Ronin bridge attack ($624M), Poly Network ($611M), and BNB ($566M) illustrate the severity [15]. Despite this, existing research often addresses only one or a few attacks while ignoring others [19].

This paper addresses the gap by performing a comprehensive threat analysis. We identify 38 attacks and vulnerabilities across five categories and provide countermeasures for each. Our contributions are:

- Identify all security threats across five classifications: core BC, Network, Interoperability, Social Engineering, and Vulnerabilities.
- Identify countermeasures for every identified threat.

The remainder of this paper is organized as follows: Section II reviews related work; Section III presents the security threats; Section IV presents countermeasures; Section V concludes.

## II. RELATED WORKS

Much work has been done on BC security, but it is scattered across components and attack types. No single work considers all threats and solutions holistically. We surveyed peer-reviewed papers from 2016–2026 addressing security in BCs and BC interoperability.

In [2], several consensus attacks are described—Sybil, 51%, Nothing at Stake, and double-spending—but without full solutions. Work in [7] proposes ETH Relay and addresses relay-specific attacks (relay poisoning, incentive attacks) with countermeasures but ignores other threat classes. A comprehensive BC vulnerability taxonomy is introduced in [15], covering many attack categories with countermeasures, but some entries lack detailed impact analysis.

In [19], BC security threats are categorized across consensus, network, and application layers with detailed solutions. Work in [23] examines 20 Bitcoin-specific attacks with descriptions, targets, consequences, and countermeasures—a model for thorough threat analysis, though it omits interoperability. In [25], communication protocol vulnerabilities in Algorand, Aptos, Avalanche, Redbelly, and Solana are studied, showing targeted load, transient failure, packet dropping, and stopping attacks.

Work in [28] provides mathematical proof showing the effectiveness of a sore loser attack defense for atomic swaps. Social engineering is addressed in [36] (phishing in BC environments) and [37] (address manipulation and homograph attacks in smart contracts). Cross-chain bridge attack detection is the focus of [41], which identifies unrestricted deposit emitting, inconsistent event parsing, and unauthorized unlocking.

While these works make important contributions, none provides a unified threat analysis spanning all five categories relevant to BC interoperability. This paper fills that gap.

## III. SECURITY THREATS

BC interoperability is composed of multiple components, each susceptible to different attacks. We classify threats into five categories: BC attacks (targeting core BC infrastructure), network attacks (abusing communication), interoperability-specific attacks (unique to cross-chain systems), social engineering (targeting users), and vulnerabilities (code and design weaknesses enabling attacks). We focus on threats applicable broadly to interoperable systems, excluding niche or deprecated attacks and privacy-specific threats.

### *A. BC Attacks*

*51% Attack:* An attacker controlling more than 51% of BC nodes or computational power can dominate consensus, enabling illegitimate blocks and further fraud [16]. Detection metrics include a single user mining a disproportionate share of blocks [17].

*Sybil Attack:* The attacker creates many pseudonymous identities to gain undue consensus influence, enabling denial-of-service by controlling message routing and reducing authentic node redundancy [20].

*Double-Spending:* The attacker spends the same cryptocurrency multiple times, often due to transaction conflicts or desynchronization in the lock-and-burn process of cross-chain transactions [16][17].

*Block Withholding Attack:* A malicious miner submits partial proof-of-work, preventing the pool from receiving rewards while remaining anonymous [19].

*Selfish Mining:* The attacker hides completed proof-of-work to gain an advantage on the next block, earning unfair rewards and potentially causing forks. Requires approximately 50% of total computational power [17][19].

*Fork After Withholding Attack:* Combines block withholding and selfish mining so only specific pool members receive increased rewards while honest miners waste resources. [19].

*Private Key Leakage:* An attacker steals a BC account's private key via network methods, physical node capture, or compromised RPC/notary nodes, gaining full account control and theft of funds [11][16].

*Nothing at Stake Attack:* The attacker replicates the entire BC history, accumulating stake in their copy, then uses the longer chain to take over—common in naive proof-of-stake BCs [21].

*Time-Jacking Attack:* The attacker plants fake-timestamped nodes to manipulate miner clocks, enabling alternate chain acceptance and double-spending [23].

*Replay Attack:* The attacker intercepts and retransmits transaction packets or exploits missing magic value or nonce validation to duplicate transactions across chains. A common consequence is double spending [6][19].

*Transaction Malleability:* The attacker modifies a transaction's unique ID without invalidating it, enabling double deposits and double withdrawals on exchanges [19][23].

*Reentrancy Attack:* The attacker exploits a publicly visible smart contract vulnerability by calling it repeatedly. Since contracts are immutable once deployed, the vulnerability cannot be patched [19].

### *B. Network Attacks*

*Denial-of-Service (DoS/DDoS):* Nodes or resources are made unavailable by flooding with invalid transactions. Larger networks with more nodes are more resilient. Bitcoin alone has experienced 142 documented DoS attacks [23].

*Botnet:* A network of compromised nodes performs attacks on behalf of an attacker, including DoS, eclipse attacks, credential brute-forcing, and unauthorized mining [24].

*Packet-Dropping Attack:* The attacker causes targeted nodes to drop packets, halting communication. TCP-based BCs (e.g., Algorand, Redbelly) can lose 95% of bandwidth; recovery is slow [25].

*Stopping Attack:* The attacker shuts down a large fraction of BC nodes—at 90% compromise, throughput approaches zero and transactions stall. Recovery is further hindered by consensus leader selection failures (e.g., Solana) [25].

*Eclipse Attack:* The attacker surrounds a target node with malicious nodes, isolating it from the legitimate network. Enables privacy loss, resource manipulation, and double-spending. Known to affect relay nodes in interoperable BCs [16][18][19].

*Partitioning Attack:* Abuses peer management to prevent communication among a set of nodes. Often performed by forcing the download of invalid blocks; closely related to eclipse attacks [13].

*Man-in-the-Middle Attack:* The attacker intercepts and potentially modifies cross-chain communication using IP, ARP, or DNS spoofing, causing privacy loss and transaction manipulation [24][26].

### *C. Interoperability Attacks*

*Relay Poisoning Attack:* The attacker submits invalid or valid-but-fraudulent block headers to a relay to reorganize the chain or approve unauthorized cross-chain transactions [7][18].

*Oracle Poisoning:* The attacker feeds false exchange rate data to the BC oracle, enabling unfair liquidation of collateral or unauthorized collateral-free unlocking. Risk is highest with single-source oracles [18].

*Goldfinger Attack:* A 51%-style attack specifically targeting a BC to destroy it for economic or political gain. Particularly dangerous for new sidechains during bootstrapping due to low initial stake [1][27].

*Sore Loser / Lockup Griefing Attack:* A participant halts communication mid-transaction, leaving assets locked indefinitely. Nearly impossible to fully prevent due to the distributed nature of BCs [28].

*Unrestricted Deposit Emitting:* The attacker bypasses asset locking by exploiting unsafe transfer functions in smart contracts, enabling deposits at the destination without a corresponding source lock (e.g., Multichain, Meter.io) [41].

*Unauthorized Unlocking:* The attacker obtains asset unlock keys through key leakage and unlocks assets without proper source-chain authorization, enabling double-spending (e.g., Anyswap) [41].

### *D. Social Engineering*

*Social Engineering:* Attackers manipulate users into exposing private keys or credentials through impersonation, pretexting, reverse social engineering, or whaling [35][36].

*Phishing:* Fake emails, websites, or wallets trick users into surrendering credentials or funds. Techniques include clone sites with lookalike domains, bloating schemes, fake ICOs, and malicious wallet apps [33][35].

*Homograph Visual Cognitive Deception:* The attacker substitutes visually similar Unicode characters in smart contract code, addresses, or function headers. Success rates of 50% have been demonstrated on hardware wallets [34][37].

### *E. Vulnerabilities*

*Smart Contract Vulnerabilities:* Once deployed, smart contracts are immutable and publicly visible, making any vulnerability permanent and exploitable indefinitely. Risks include unprotected self-destruct functions and business logic flaws, which can lead to theft of funds and other attacks [6][15].

*Public Key Cryptography Vulnerabilities:* Fault attacks, voltage/clock glitches, and related-key attacks can compromise RSA, ECC, ECDSA, EdDSA, and Schnorr keys, particularly on physical wallet hardware [6][29][30][31][32].

*Improper Value Validation:* Missing magic value or nonce validation in smart contracts allows replay attacks across chains [6].

*Smart Contract Visibility Vulnerability:* Functions without access labels default to public, exposing data and allowing unauthorized calls [6].

*Weak Randomness:* Using block hash, number, or timestamp for randomness is predictable and exploitable (e.g., SmartBillions lottery theft) [6].

*Transaction Order Dependence:* Higher-fee transactions are prioritized, enabling an attacker to copy and front-run another user's solution to claim rewards [6].

*Asset Counterfeiting:* Cryptocurrency issued without proper collateralization proof, causing inflation and trust loss. Particularly affects Zcash-based BCs [18].

*Verifier's Dilemma:* Validator nodes cease verifying transactions due to low perceived risk, high cost, or assumed redundancy, enabling fraudulent transactions [39].

*Invalid Consensus Proofs:* Attackers submit fake blocks with forged sync committee signatures or replay valid proofs to duplicate transactions. Enables double-spending [40].

*Inconsistent Event Parsing:* The event parser incorrectly validates malicious cross-chain events or miscounts assets, causing abusable unexpected behavior (e.g., THORChain, pNetwork) [41].

### F. Summary

Table I summarizes all 38 identified attacks and vulnerabilities, their categories, impacts, and references.

| Attack | Category | Impact | Refs |
|---|---|---|---|
| 51% Attack | BC | Attacker controls consensus, fraud | [15-17] |
| Sybil Attack | BC | Consensus disruption, DoS | [16][17][20] |
| Double-Spending | BC | Fraudulent spending | [16][17] |
| Block Withholding | BC | Disrupted mining, reward loss | [19] |
| Selfish Mining | BC | Unfair rewards, potential fork | [13][17][19] |
| Fork After Withholding | BC | Forking, reward theft | [13][19] |
| Private Key Leakage | BC | Account takeover, fund loss | [11][16] |
| Nothing at Stake | BC | BC hijack | [2][21] |
| Time-Jacking | BC | Double-spending, resource waste | [8][23] |
| Replay Attack | BC | Transaction duplication | [6][19] |
| Tx Malleability | BC | Double-spending, fund theft | [19][23] |
| Reentrancy | BC | Financial exploitation | [19] |
| DoS / DDoS | Network | Nodes offline, BC halted | [13][16][17] |
| Botnet | Network | DoS, eclipse, further attacks | [23][24] |
| Packet-Dropping | Network | Communication halted | [25] |
| Stopping Attack | Network | BC halted, lost transactions | [25] |
| Eclipse Attack | Network | Privacy loss, double-spending | [16][18][19] |
| Partitioning | Network | Eclipse attack | [13] |
| Man-in-the-Middle | Network | Privacy loss, tx manipulation | [10][24][26] |
| Relay Poisoning | Interop. | Fraudulent cross-chain txns | [7][18] |
| Oracle Poisoning | Interop. | Fund loss, unfair exchange | [9][18] |
| Goldfinger | Interop. | Fork, double-spending | [1][23][27] |
| Sore Loser / Lockup Griefing | Interop. | Assets locked indefinitely | [1][28] |
| Unrest. Deposit Emit. | Interop. | Double-spending | [41] |
| Unauthorized Unlock | Interop. | Double-spending | [41] |
| Social Engineering | Social Eng. | Account/fund loss | [19][35-37] |
| Phishing | Social Eng. | Account/fund loss | [23][33][35] |
| Homograph Deception | Social Eng. | Tx alteration, fund theft | [34][37] |
| Smart Contract Vulns | Vulnerab. | Fund loss, attacks | [1][6][15] |
| Public Key Crypto Vulns | Vulnerab. | Key/account compromise | [6][29-32] |
| Improper Value Validation | Vulnerab. | Replay attacks | [6] |
| Visibility Vulnerability | Vulnerab. | Unauthorized access | [6] |
| Weak Randomness | Vulnerab. | Exploitable randomness | [6] |
| Tx Order Dependence | Vulnerab. | Reward theft | [6] |
| Asset Counterfeiting | Vulnerab. | Trust loss, inflation | [15][18] |
| Verifier's Dilemma | Vulnerab. | Fraudulent txns | [15][39] |
| Invalid Consensus Proofs | Vulnerab. | Double-spending | [15][40] |
| Inconsistent Event Parse | Vulnerab. | Unexpected behavior | [15][41] |

**TABLE I. BC INTEROPERABILITY ATTACKS AND VULNERABILITIES**

## IV. COUNTERMEASURES

We now present countermeasures for each threat, organized by category. Common patterns emerge across categories: monitoring and detection systems address 11+ threats; validation techniques counter replay, consensus proof, and event parsing attacks; timestamp mechanisms defend against replay, time-jacking, and double-spending; reputation systems deter Sybil, eclipse, and Goldfinger attacks; and code audits eliminate many smart contract vulnerabilities before deployment.

### *A. BC Attack Countermeasures*

*51% Attack:* Increase overall network hash rate; distribute nodes among many unique entities; adopt Proof-of-Stake or Proof-of-Authority consensus; increase PoS validator count to reduce bribery opportunities; monitor for suspicious mining patterns [12][16][19][23][42].

Sybil Attack*:* Use permissioned or hybrid BCs where all participants are known; or implement a reputation system assigning trust scores based on age and behavior, preventing rapid multi-account creation from gaining consensus power [2][20].

*Double-Spending:* Combine higher hash rate, transaction validity checks (hash, timestamp, nonce), lock-and-burn asset transfers with post-burn delay, timestamp logging, and notification blocking to prevent repeat offenders [17][23][35][42][43].

*Block Withholding:* Deploy monitoring with verifiable shares (Stratum protocol); use reputation-based reward reduction; apply revenue distribution adjustments; limit pool participation to trusted users; use zero-knowledge proofs for full proof-of-work commitment [19][23].

*Selfish Mining:* Optimize block propagation; apply dynamic difficulty adjustment; use protocol upgrades to close known exploits; introduce randomness in fork selection; use unforgeable timestamps to prioritize freshly published blocks [19][23].

*Fork After Withholding:* Use improved block confirmation to detect competing forks; enhance consensus to reject malicious forks; employ silent timestamps; apply incentive systems rewarding complete proof-of-work [19].

*Private Key Leakage:* Limit wallet balances; encrypt private keys with password-protected symmetric keys; require multi-signature transactions, distributing key storage across separate locations [11].

*Nothing at Stake:* Apply a checkpoint latch that prevents BC reorganization once a branch reaches a threshold size tied to validator liquidation periods [21].

*Time-Jacking:* Enforce tighter transaction acceptance time windows; use system time as a fallback; obtain time samples from trusted peers via NTP [23].

*Replay Attack:* Validate magic values and nonces for all transactions; use unique zero-knowledge-proof-backed locks in atomic swaps; enforce timestamp-based validity windows [6][18][19][24][22].

*Transaction Malleability:* Implement Segregated Witness (also improves scalability and atomic swap performance); enforce strict signature encoding; modify transaction ID generation to exclude mutable fields; use collisionless hash algorithms [19][23].

*Reentrancy:* Apply Check-Effects-Interactions pattern; use mutex locks/state variables for synchronization; audit code for recursive call vulnerabilities; implement static/dynamic analysis tools and fuzz testing [15][19].

### *B. Network Attack Countermeasures*

*DoS/DDoS:* Increase peer connections; restrict access to trusted IPs; throttle transaction rates; block known attack patterns; adopt Proof-of-Activity consensus; deploy fast-verification authentication; use ML-based traffic monitoring with isolated containment [11][17][23].

*Botnet:* Require node authentication; educate users to avoid malicious files and links; monitor for ongoing attack signatures; verify node validity continuously before each transaction [24][5].

*Packet-Dropping:* Switch to QUIC protocol (resilient over TCP); add confirmation messages; use Diffie-Hellman key exchange written to the BC for tamper-evident key negotiation [25][44].

*Stopping Attack:* Increase transaction pool size; disable node throttling; modify lockout configuration to allow any node to lead after mass failures and restarts [25].

*Eclipse Attack:* Ensure at least one honest node (raises attack cost to 51% attack level); use random peering; diversify connections geographically; maintain reputation-based trust scores; deploy monitoring with ML-based anomaly detection; maintain critical node backups [18][19][23].

*Partitioning Attack:* Add more BC nodes; allow concurrent multi-block downloads (minimum three sessions eliminate most partition vectors); migrate from TCP to QUIC [13][25].

*Man-in-the-Middle:* Deploy OAuth 2.0 with multi-factor authentication; issue digital certificates to all nodes via trusted providers (Amazon, Upvest, Kaleido); use public-private key encryption for all cross-chain messages [10][24][26].

### *C. Interoperability Countermeasures*

*Relay Poisoning:* Validate both block headers and transactions; require only one honest node to dispute invalid submissions; wait additional confirmation blocks before finalization to prevent chain reorganization [7][18].

*Oracle Poisoning:* Aggregate data from multiple independent sources and oracles; implement a dispute mechanism for oracle discrepancies; provide financial incentives for accurate reporting [18].

*Goldfinger Attack:* Apply merged-staking to protect new sidechains during bootstrapping; deploy monitoring with double-spending detection; reputation system to enforce proper behavior; limit mining pool size [1][23].

*Sore Loser / Lockup Griefing:* Add a premium payment as a penalty for abandonment (refundable on completion); implement a premium distribution phase in HTLC atomic swap protocols; shorten lock periods via immediate post-transaction unlock [1][28].

*Unrestricted Deposit Emitting:* Require verified lock events before any deposit is permitted; use only secure transfer functions; deploy automated pattern-matching detection tools [41].

*Unauthorized Unlocking:* Apply private key leakage protections; restrict unlock access to authorized cross-chain verification nodes only; deploy detection tools monitoring unlock transaction signatures [41].

### *D. Social Engineering Countermeasures*

*Social Engineering:* Enforce dual control for transactions; educate users on malware, suspicious links, communication tones, and unknown accounts; avoid contracts permitting address changes or string comparisons; deploy signature-based detection systems; build attack-tolerant systems with logging [35][36][37][38].

*Phishing:* Use browser anti-phishing extensions; train users to verify URLs, certificates, and email details; deploy URL and content detection algorithms; require two-factor authentication [23][33][35][36].

*Homograph Deception:* Educate users to verify full addresses and checksums; deploy automated monitoring for known attack patterns; ensure smart contracts include fallback functions; verify cross-contract function selectors via hex viewing [34][37].

### *E. Vulnerability Countermeasures*

*Smart Contract Vulnerabilities:* Follow secure development practices; conduct pre-deployment code reviews, strict testing, and formal modeling; avoid self-destruct functions; use vulnerability scanners; offer bug bounties; provide developer security education [1][6][15].

*Public Key Cryptography:* Use vetted libraries (e.g., mbedTLS); implement check values against fault attacks; generate keys with validated RNGs (XEdDSA, VXEdDSA); apply multi-signature redundancy; enhance algorithm designs (Schnorr/DSA with re-keyed hash; ElGamal with Gaussian integers) [6][29][30][31][32].

*Improper Value Validation:* Enforce magic value and nonce validation for every cross-chain transaction [6].

*Visibility Vulnerability:* Explicitly label all functions and variables; audit access controls; encrypt sensitive data [6].

*Weak Randomness:* Replace insecure sources with secure RNG libraries designed for smart contracts (RANDAO, RBGC) [6].

*Transaction Order Dependence:* Use a commit-reveal commitment scheme so solutions are hidden until processed, preventing front-running [6][3].

*Asset Counterfeiting:* Require collaterals with proof-of-balance checks; embed anti-counterfeiting rules in consensus; mandate distributed signatures for unknown-party transactions [15][18].

*Verifier's Dilemma:* Deploy a canary protocol that submits harmless invalid transactions to test vigilance; require nodes to stake on approvals, thereby incentivizing careful validation [39].

*Invalid Consensus Proofs:* Verify sync committee authenticity and block signatures; check block height to reject repeated proof submissions; apply DoS defenses to resist consensus proof flooding [15][40].

*Inconsistent Event Parsing:* Redesign parsers to include verification before any cross-chain asset issuance; deploy real-time detection tools to identify parsing attack signatures [15][41].

### *F. Summary*

Table II summarizes all countermeasures by attack and category.

| Attack | Category | Key Countermeasures | Refs |
|---|---|---|---|
| 51% Attack | BC | Higher hash rate, more nodes, PoS validators, monitoring | [12][16][19][23][42] |
| Sybil Attack | BC | Permissioned BC, reputation system | [2][20] |
| Double-Spending | BC | Hash rate, block validation, lock-and-burn, timestamps | [17][23][27][35][42][43] |
| Block Withholding | BC | Monitoring, reputation, Stratum protocol, credit mechanism | [19][23] |
| Selfish Mining | BC | Block propagation opt., dynamic difficulty, timestamps | [19][23] |
| Fork After Withholding | BC | Improved block confirmation, consensus enhancement | [19][23] |
| Private Key Leakage | BC | Limit wallet funds, encrypt key, multi-sig transactions | [11] |
| Nothing at Stake | BC | Checkpoint latch | [21] |
| Time-Jacking | BC | Tighter time restrictions, NTP, trusted peers | [23] |
| Replay Attack | BC | Magic value & nonce validation, unique locks, timestamps | [6][18][19][24][22] |
| Tx Malleability | BC | Segregated Witness, strict sig encoding, validation | [19][23] |
| Reentrancy | BC | Check-Effects-Interactions, mutex locks, audits | [15][19] |
| DoS / DDoS | Network | More nodes, throttling, PoA consensus, monitoring | [11][17][23] |
| Botnet | Network | Node authentication, user education, monitoring | [24][5] |
| Packet-Dropping | Network | QUIC protocol, confirmation mechanism, DH key exchange | [25][44] |
| Stopping Attack | Network | Larger tx pool, disable throttling, lockout config | [25] |
| Eclipse Attack | Network | Random peering, reputation system, diverse connections | [18][19][23] |
| Partitioning | Network | More nodes, concurrent downloads, QUIC protocol | [13][25] |
| Man-in-the-Middle | Network | Authentication, digital certificates, encryption | [10][24][26] |
| Relay Poisoning | Interop. | Header &tx validation, extra confirmations | [7][18] |
| Oracle Poisoning | Interop. | Multiple oracles/sources, dispute mechanism, incentives | [18] |
| Goldfinger | Interop. | Merged-staking, monitoring, reputation system, limit mining pool size | [1][23] |
| Sore Loser / Lockup | Interop. | Premiums, HTLC premium phase, incentives | [1][28] |
| Unrest. Deposit Emit. | Interop. | Lock verification, secure transfer functions, detection | [41] |
| Unauthorized Unlock | Interop. | Key protection, access controls, detection tool | [41] |
| Social Engineering | Social Eng. | Dual control, user education, detection system | [35][36][37][38] |
| Phishing | Social Eng. | Anti-phishing tools, user education, 2FA | [23][33][35][36] |
| Homograph Deception | Social Eng. | Checksum use, monitoring, verify selectors | [34][37] |
| Smart Contract Vulns | Vulnerab. | Best practices, testing, audits, formal modeling | [1][6][15] |
| PK Crypto Vulns | Vulnerab. | Secure library, check values, strong RNG, redundancy | [6][29-32] |
| Improper Validation | Vulnerab. | Validate magic values and nonces | [6] |
| Visibility Vuln. | Vulnerab. | Label all functions/data, audit, encrypt data | [6] |
| Weak Randomness | Vulnerab. | Secure RNG (RANDAO, RBGC) | [6] |
| Tx Order Dependence | Vulnerab. | Commit-reveal scheme | [6][3] |
| Asset Counterfeiting | Vulnerab. | Collateral, consensus rules, distributed signatures | [15][18] |
| Verifier's Dilemma | Vulnerab. | Canary protocol, stake-based verification requirement | [39] |
| Invalid Consensus Proofs | Vulnerab. | Verify sync committee, block height checks | [15][40] |
| Inconsistent Event Parse | Vulnerab. | Fix parser design, automated detection tool | [15][41] |

**TABLE II. ATTACK COUNTERMEASURES**

## V. CONCLUSION AND FUTURE WORK

Effective BC interoperability requires comprehensive security from all threat vectors—not merely those specific to cross-chain protocols. BC attacks such as double-spending, replay, transaction malleability, and reentrancy can cause cross-chain fraud with asset loss. Network-layer attacks, including DoS and man-in-the-middle, can halt communication entirely. Interoperability-specific threats such as unrestricted deposit emitting, unauthorized unlocking, and relay/oracle poisoning can bypass bridge mechanisms entirely. Social engineering exploits the human layer even when technical security is strong. Finally, smart contract code and public key cryptography vulnerabilities can be exploited indefinitely once deployed.

Several cross-cutting countermeasure patterns emerged: monitoring and detection systems address over 11 distinct threats; validation techniques prevent replay and consensus attacks; timestamp mechanisms defend against time-based attacks; reputation systems deter Sybil, eclipse, and Goldfinger attacks; and code audits prevent a wide range of smart contract vulnerabilities before deployment.

However, security evolves continuously. Countermeasures must be combined into layered defenses capable of handling both known and novel threats. Privacy in BC interoperability—currently under-addressed—will be the focus of our subsequent work, followed by the development of enhanced architectural patterns for secure, attack-resistant BC interoperability bridges.